\documentclass[sigconf, dvipsnames, authorversion]{acmart}
\usepackage{float}
\usepackage[T1]{fontenc}
\usepackage[utf8]{inputenc}

\restylefloat{table}
\usepackage{enumitem}
\usepackage{array}
\usepackage{longtable}
\usepackage{algorithm}
\usepackage{bbm}
\usepackage[noend]{algpseudocode}
\usepackage{float}
\usepackage{tikz}
\usepackage{subcaption}
\usepackage{enumitem}
\usepackage{tikz}
\usepackage[utf8]{inputenc}
\usepackage{pgfplots}
\pgfplotsset{compat=newest}
\usepgfplotslibrary{groupplots}
\usepgfplotslibrary{dateplot}
\usetikzlibrary{bayesnet}
\usepackage{color,colortbl}
\definecolor{mygray}{gray}{0.6}
\newcommand{\PreserveBackslash}[1]{\let\temp=\\#1\let\\=\temp}
\newcolumntype{C}[1]{>{\PreserveBackslash\centering}p{#1}}
\newcolumntype{R}[1]{>{\PreserveBackslash\raggedleft}p{#1}}
\newcolumntype{L}[1]{>{\PreserveBackslash\raggedright}p{#1}}

\AtBeginDocument{%
  \providecommand\BibTeX{{%
    \normalfont B\kern-0.5em{\scshape i\kern-0.25em b}\kern-0.8em\TeX}}}

\newcommand{\ranking}[0]{\pi}
\newcommand{\doc}[0]{d}

\newcommand{\groups}[0]{\mathcal{G}}
\newcommand{\group}[0]{g}
\newcommand{\groupA}[0]{A}
\newcommand{\groupB}[0]{B}

\newcommand{\query}{q}

\newcommand{\docsAll}[0]{\mathcal{D}}
\newcommand{\docs}[1]{\mathcal{D}_{#1}}
\newcommand{\docsA}[0]{\docs{\groupA}}
\newcommand{\docsB}[0]{\docs{\groupB}}

\newcommand{\targetExposure}[1]{\mathbf{P}_{#1}}

\newcommand{\treatmentExposure}[1]{\mathbf{\tilde{P}}_{#1}}

\newcommand{\exposureSumTarget}[1]{\mathbf{E}_{#1}}
\newcommand{\exposureSum}[1]{\mathbf{\tilde{E}}_{#1}}

\newcommand{\numdocs}[0]{k}

\newcommand{\topk}[2]{#1_{\leq #2}}

\newcommand{\metric}[0]{\Delta}
\newcommand{\metricAbs}[0]{\metric_{\text{abs}}}
\newcommand{\metricSq}[0]{\metric_{\text{sq}}}
\newcommand{\metricKL}[0]{\metric_{\text{KL}}}
\newcommand{\metricDiff}[0]{\metric_{\text{diff}}}

\newcommand{\patience}[0]{\gamma}

\newcommand{\indicator}[1]{\mathbb{I}\left(#1\right)}
\setcopyright{iw3c2w3}
\begin{document}
\copyrightyear{2021}
\acmYear{2021}
\acmConference[WWW '21]{Proceedings of the Web Conference 2021}{April 19--23, 2021}{Ljubljana, Slovenia}
\acmBooktitle{Proceedings of the Web Conference 2021 (WWW '21), April 19--23, 2021, Ljubljana, Slovenia}

\acmPrice{}
\acmDOI{10.1145/3442381.3450080}

\acmISBN{978-1-4503-8312-7/21/04}

\title{Estimation of Fair Ranking Metrics with Incomplete Judgments}


\author{\"Omer K{\i}rnap}
\affiliation{
   \institution{University College London}
   \city{London}
   \country{UK}
}
\email{omer.kirnap.18@ucl.ac.uk}

\author{Fernando Diaz}
\authornote{Now at Google.}
\affiliation{
   \institution{Microsoft Research}
   \city{Montreal}
   \state{Quebec}
   \country{Canada}
}
\email{diazf@acm.org}

\author{Asia Biega}
\authornote{Work in part done while at Microsoft Research.}
\affiliation{
   \institution{Max Planck Institute \\for Security and Privacy}
    \city{Bochum}
   \country{Germany}
}
\email{asia.biega@acm.com}

\author{Michael Ekstrand}
\affiliation{
   \institution{People \& Information Research Team\\Boise State University}
   \city{Boise}
   \state{Idaho}
   \country{USA}
}
\email{michaelekstrand@boisestate.edu}

\author{Ben Carterette}
\affiliation{
   \institution{Spotify Research}
   \city{New York}
   \country{USA}
}
\email{benjaminc@spotify.com}

\author{Emine Y{\i}lmaz}
\affiliation{
   \institution{University College London}
   \city{London}
   \country{UK}
}
\email{emine.yilmaz@ucl.ac.uk}



\begin{abstract}
There is increasing attention to evaluating the \textit{fairness} of search system ranking decisions.  
These metrics often consider the membership of items to particular groups, often identified using \emph{protected attributes} such as gender or ethnicity.
To date, these metrics typically assume the availability and completeness of protected attribute labels of items.
However, the protected attributes of individuals are rarely present, limiting the application of fair ranking metrics in large scale systems. 
In order to address this problem, we propose a sampling strategy and estimation technique for four fair ranking metrics. We formulate a robust and unbiased estimator which can operate even with very limited number of labeled items.  
We evaluate our approach using both simulated and real world data. 
Our experimental results demonstrate that our method can estimate this family of fair ranking metrics and provides a robust, reliable alternative to exhaustive or random data annotation. 
\end{abstract}


\keywords{information retrieval, evaluation, fairness, fair ranking}


\maketitle

\section{Introduction}
\label{sec:introduction}

Information retrieval (IR) evaluation often focuses on the effectiveness of a system, but there is also a significant history of measuring additional properties of a system's output or behavior, such as novelty and diversity \citep{Clarke2008-fx}.
In the last several years, there has been increased interest in the \emph{fairness} of information retrieval systems \citep{Roegiest2019-ks}, with a number of metrics being proposed \citep{beutel2019fairness,diaz2020evaluating, zehlike2017fa, biega2018equity, singh2018fairness, yang2017measuring}.
While fairness metrics and constructs come in different flavors and aim at different goals, they all attempt to measure the \emph{societal} impacts of the system decisions, and in particular to ensure that those impacts are equitably distributed.

In this paper, we study metrics for \emph{provider group fairness}.
This means that the fairness goal is to ensure that the \emph{providers} of items or documents are treated fairly (as opposed to consumers or other stakeholders in multi-sided information access \citep{Burke2017-ne}). 
One way to evaluate this is by measuring whether the exposure different providers receive from the system is equitably distributed among providers of documents with comparable relevance \citep{diaz2020evaluating, biega2018equity}.
In this spirit, one class of metrics seeks to ensure that socially-salient \emph{groups} of providers receive comparable exposure; that is, to measure whether providers of, for example, a particular gender, ethnicity, professional seniority, or other group potentially subject to discrimination are systematically disadvantaged in the system's results.
This can be measured by aggregating exposure over provider groups, or by measuring the representation of provider groups in result lists \citep{yang2017measuring}.

These metrics require the availability of group membership annotations: in order to determine if system results are unfairly discriminating against particular groups, we need to know which groups the various entities in its corpus are associated with.
These annotations are often difficult to acquire; reasons for this difficulty include general unavailability, legal and/or ethical restrictions on its collection or use (particularly since group membership is often sensitive personal information), or the cost of expert annotation to e.g. identify content creators' group identities from publicly-available data. These challenges are reflected in analyses of the needs of practitioners building fair systems.
In a recent survey of machine learning practitioners by \citet{holstein2019improving}, practical approaches to auditing fairness with limited data were mentioned as one of the most pressing issues. 
To address the needs, recent work has looked at mechanisms for auditing \citep{kallus2019assessing} and optimizing \citep{hashimoto2018fairness,lahoti2020fairness}  systems in the absence of such labeled data, with some success but also notable limitations, and none of this work has yet been applied to ranking, retrieval, or recommendation systems.

In this work, we consider the case where group membership annotations are available, but at a cost, so it is not practical to obtain complete labels for the underlying data but a strategically-selected sample can be labeled.
This is the case, for instance, when annotations must be provided by human annotators, and researchers or system maintainers wish to make effective use of a limited budget for hiring annotators. 

Our goal is to develop \emph{statistical estimation techniques} that enable accurate estimation of a provider group fairness metric, applied to an IR system's ranked outputs, by acquiring group membership annotations for a sample of documents in its corpus.
Inspired by the work by Pavlu \textit{et al.}~\cite{pavlu2007practical} on estimating information retrieval effectiveness metrics using incomplete judgments, along with other work in this line~\cite{yilmaz2006estimating, yilmaz2008simple, aslam2006statistical, CarteretteMinimal, BuckleyIncomplete, roitero2020effectiveness, roitero2018reproduce}, we show how unbiased estimates of various fairness metrics can be computed using estimators based on the Horvitz-Thompson estimator~\cite{Thompson02}. 
To our knowledge, this is the first application of information retrieval metric estimation from incomplete data to the problem of assessing system fairness.
In particular, we show how unbiased estimates of a few proportion-based metrics ~\cite{yang2017measuring} and an exposure based metric\cite{diaz2020evaluating} can be approximated with a significantly smaller number of group membership annotations. While we focus on these particular metrics in this paper, the techniques can easily be extended to estimate other fairness metrics when group membership annotations are incomplete.

\section{Related Work}
\label{sec:related}
In this section, we present the connection between  information retrieval (IR) evaluation techniques and the fairness of IR systems. 
\subsection{Evaluating Information Retrieval}

IR systems find information believed to match a user's information need from a \emph{corpus} of \emph{documents} (or other items).
In their most common configuration, which we study here, they return a ranked list of \emph{documents} in response to a \emph{query} (for a search task) or a user's context and interest profile (for a recommendation task).

The performance of these systems is often evaluated with a variant of the ``Cranfield protocol'' \cite{voorhees2001philosophy}, where the system's rankings are compared with a set of ground-truth \emph{relevance judgements} and its accuracy measured using a metric such normalized discounted cumulative gain (nDCG) \cite{jarvelin2002ndcg} or expected reciprocal rank \cite{chapelle2009err}.
nDCG and related metrics assess the system's ability to place the most relevant documents at the top of its ranked result lists. They also prioritize accuracy at the top of the list, because users tend to pay more attention to the first few results.

The relevance judgments come from variety of sources, depending on the experimental setting.
In some cases, they are provided by expert assessors; in others, they are derived from user signals such as clicks, purchases, and ratings.
Most metrics, in their naive forms, assume that relevance data is complete: that the evaluation not only know the relevance of every ranked document, but also the relevance of every document in the corpus so it can penalize a system for failing to retrieve relevant documents.

Sampling techniques estimate IR metrics with with incomplete judgments \citep{aslam2006statistical, yilmaz2006estimating, yilmaz2008simple}.
These methods assume that relevance \emph{can} be assessed for any document with respect to a particular query, but at a cost; they approach the problem by strategically selecting documents to assess in order to accurately estimate the metric based on annotations for a subset of the corpus.
We extend these ideas to assess the fairness of a system's rankings instead of its performance; this presents both new opportunities (since a document's protected group affiliation is independent of any particular query) and challenges (methods exploiting the structure of a performance metrics to improve sampling efficiency do not directly translate to fairness metrics).

\subsection{Fairness in Algorithmic Systems}

Algorithmic fairness is a broad field with many interlocking concepts and sometimes competing; \citet{Mitchell2018-jt} provide a useful overview.
Most of this work, however, has focused on classification and regression models.

One key distinction in the algorithmic fairness literature is the line between \emph{individual} and \emph{group} fairness \citep{dwork2012fairness}.
Individual fairness is concerned with ensuring that similar individuals receive similar decisions; in an IR system, this could mean that documents with similar relevance to a query should have equivalent probabilities of being retrieved in a valuable ranking position \citep{biega2018equity}.
Group fairness, as we discussed in the introduction, ensures that data subjects don't experience disparate treatment, decision outcomes, or error on the basis of their group identity or membership.
This is often operationalized through the concepts of \emph{protected groups} and \emph{sensitive attributes}, inspired by U.S. anti-discrimination law, resulting in fairness objectives such as ``equality of opportunity'', the constraint that system decisions should be conditionally independent of group membership given true outcomes \citep{hardt2016equality}.

An additional distinction that is particularly relevant to information retrieval systems is the difference between ensuring fairness for information providers, consumers, and other stakeholders in multi-sided systems \citep{Burke2017-ne}.

\subsection{Fairness of Ranked Outputs}
Ranking systems, including search, recommendation, and matchmaking, have dynamics that are different than the classification and regression models typical of algorithmic fairness research.
This flows from two interconnected problems: first, rank positions are exclusive, in that only one document can be in the first position of a given set of search results; second, such systems often exhibit \emph{position bias}, where users are more likely to inspect results higher in the ranking~\cite{chuklin2015click, craswell2008experimental}. Documents that are ranked on top receive higher click rates even if they are not actually relevant to a query~\cite{joachims2007search}.
Broadly, there are two families of methods used for measuring the fairness of ranking systems:

\textbf{Exposure Based Methods}. Exposure can be defined as user's discovery of different documents in a ranked list. In other words, it is kind of the distribution of user's attention to documents in ranked list. Exposure-based metrics can quantify the level of attention discrepancy spent in some documents on top but not the lower ranks. In the context of amortized evaluation, Biega \textit{et al.}~\cite{biega2018equity} studies equity of attention among positions in rankings that are relevant to a query. Morik \textit{et al.}~\cite{morik2020controlling} introduces a dynamic ranking scheme that optimizes the exposure metric introduced in~\cite{biega2018equity}.  Diaz \textit{et al.}~\cite{diaz2020evaluating} extend \cite{biega2018equity} to 
the context of stochastic ranking, including both individual and group fairness definitions. 
Pairwise rank fairness \citep{beutel2019fairness,kuhlman2019fare} does not directly measure exposure, but is related in that it measures the system's propensity to correctly or incorrectly rank relevant documents above irrelevant ones based on the relevant document's protected attribute: a system that is systematically more likely to correctly surface documents from the majority group than the protected group is deemed unfair.

\textbf{Proportion Based Methods}. One other framework in algorithmic fairness dictates that all groups in data should be treated similarly~\cite{pedreschi2009measuring}. This criterion has been met via statistical parity in fairness settings. Yang and Stoyanovic~\cite{yang2017measuring} propose a family of fraction based measures by comparing the distributions of different groups to adapt statistical parity into ranked outputs. Zehlike et al.~\cite{zehlike2017fa} introduces the notion of following similar distribution of corpus with every position in ranking. They systematically check whether the distribution is preserved or not in each rank.

\section{Fair Ranking Metrics}
\label{sec:prob}

In this section, we will summarize a broad family of fair ranking metrics.  
Given a query (for IR) or context (for recommendation),  assume a system ranks items from an underlying corpus $\docsAll$
producing ranking $\ranking$. A document ranked at position $i$ is denoted as $\ranking_i$; the set of top $\numdocs$ documents is denoted as $\topk{\ranking}{\numdocs}$. 
Let $\groups$ be the set of group labels and $\docs{\group}\subseteq\docsAll$ be the subset of documents with group label $\group\in\groups$.  
We consider the situation where there are two groups which partition the corpus (i.e. $\groups=\{\groupA, \groupB\}$ and $\{\docsA,\docsB\}$ is a partition of $\docsAll$).

Given a ranking $\ranking$, a group-based fair ranking metric is composed of three computations: (a) measuring the group representation in $\ranking$, (b) defining a target group representation for that query, and (c) comparing the group representation in $\ranking$ with the target group representation.

\label{subsec:metrics}

\subsection{Measuring Group Representation}
Group representation can be computed as either the proportion of the groups in the top of the ranking or the probability of examination of groups in the ranking.  

\textit{Proportion-based representation}~\cite{celis2017ranking,yang2017measuring,zehlike2017fa} measures the proportion of items belonging to different groups present in the top $k$  positions of $\ranking$. Proportion of group $\group$ in the ranking $\ranking$ can be computed as:
\begin{align}
     \treatmentExposure{\group} = \frac{|\topk{\ranking}{\numdocs} \cap \docs{\group}|}{\numdocs}
      \label{eq:disc_sum}
\end{align}

Throughout this paper, we use $k=30$ for all metrics that depend on the top $k$ portion of the ranking.

\textit{Exposure-based representation} measures the allocation of attention of searchers to items belonging to different groups~\cite{biega2018equity,diaz2020evaluating,singh2018fairness}.
Exposure is generally assumed to exponentially decrease with rank, albeit the exact formulations have differed in prior work~\cite{diaz2020evaluating}. There are some scenarios in IR tasks where the fraction of a particular group is the same for all systems. Thus, we need to pick an exposure based metric including individual ranks of documents to measure the fairness.
In this paper, we adopt a discounted exposure metric inspired by Rank Biased Precision (RBP)~\cite{moffat2008rank} and Expected Exposure in~\cite{diaz2020evaluating}. For the protected group $\group$, the following equation denotes the exposure of $\group$ in the top-k ranking results:
\begin{align}\label{eq:exposure_sum}
      \exposureSum{\group} & = (1-\patience) \sum_{i=1}^{\numdocs} \patience^{(i-1)} \indicator{\ranking_i\in\docs{\group}}
\end{align}

Here the discount factor, $\patience$, is a decay parameter controlling the importance of higher ranks. We term this metric Exposure, and use this to measure the exposure of protected group.

\subsection{Defining a Representation Target}~\label{subsec:definiton_target}
The \textit{representation target} refers to the ideal representation (i.e. proportion or distribution of exposure).  The choice of representation target depends on the application domain.  In this paper, we consider three targets often used in the literature,
\begin{itemize}
\item \textbf{Parity:} where the resource allocation should be equal for each group.
\item \textbf{Proportionality to the corpus presence:} where the resource allocation should be proportional to the number of items in the corpus that belong to a given group.
\item \textbf{Proportionality to the relevance:} where the resource allocation should be proportional to the number of items belonging to a given group that are relevant to the ranking query.
\end{itemize}
We use the notation $\targetExposure{\group}$ to refer to target proportion. We do not use a target for the Exposure metric, i.e. $\exposureSumTarget{\group}$; instead we report Equation~\ref{eq:exposure_sum} as protected group's exposure and use this as Exposure metric, a divergence from~\cite{diaz2020evaluating} that keeps with the normative origin of our proportion-based metrics, focusing on the representation of protected group members in the ranking while leaving document relevance as a separate concern.

\subsection{Divergence-Based Fairness Metrics}
\label{subsec:disc_met}
Fairness measures compare the system's proportion or exposure of a protected group with the ideal proportion or exposure suggested by the selected representation target. In this paper, we consider four divergence measures.  For proportion-based representations, these are defined as,
\begin{align}~\label{eq:divergence_metrics}
   \metricDiff & = \sum_{\group \in \groups} (\targetExposure{\group}  - \treatmentExposure{\group}) &\text{Difference}\\
   \metricAbs & = \sum_{\group \in \groups} |\targetExposure{\group}  - \treatmentExposure{\group}| &\text{Absolute Difference}\\~\label{eq_metric_sq}
    \metricSq & =\sum_{\group \in \groups} \left(\targetExposure{\group}  - \treatmentExposure{\group}\right)^2  &\text{Squared Difference}\\
     \metricKL &= \sum_{\group \in \groups} \targetExposure{\group} \log \left(\frac{\targetExposure{\group}}{\treatmentExposure{\group}}\right)  &\text{KL Divergence}
\end{align}
Definitions for exposure-based representations follow analogously.  As with the representation target, the choice of divergence measure depends on the application context of the search system.

\section{Problem Definition}
\label{sec:problemdefinition}
Given a ranking $\ranking$ and a fixed annotation budget for obtaining group labels, our goal is to develop a sampling based method that can be used to produce an unbiased estimate of the actual value of a fairness metric $\Delta$. That is, we would like to efficiently select only a subset of items in the corpus to be annotated for membership, and use those to estimate the metric of interest.

\section{Estimation Methodology}
\label{sec:method}

There are various ways in which the need for annotations could be reduced such as techniques based on active learning~\cite{CarteretteMinimal}. However, for most of these techniques, there are no guarantees that the values of metrics computed using these techniques will be unbiased estimates for the actual metric value. 

Our proposed approach for computing unbiased estimates of fairness metrics with incomplete judgments is based on the statistical estimation framework developed by Pavlu et. al.~\cite{pavlu2007practical}, which was originally proposed for reducing annotation effort in context of IR evaluation.

In this section, we first describe the sampling strategy we use, and show how unbiased estimates of fairness metrics can be computed using the sampled documents.

\subsection{Sampling}

The first step in our statistical estimation approach based on Pavlu et al.~\cite{pavlu2007practical} is to select which items should be annotated, which will be done using sampling. One of the advantages of the statistical estimation technique we use in this paper (described in the next section) is that it can be applied to compute unbiased estimates of metrics regardless of what sampling distribution is used in the sampling stage. However, the particular sampling distribution used could have an impact on the variance of the estimator, which would affect the confidence of the final estimator.

There are many different potential sampling distributions that can be used in the sampling process. Which sampling distribution to use could depend on the quantity that needs to be estimated as the estimation could be made more efficient by adopting a sampling distribution that is ideal for the fairness metric in which we are interested. For instance, if the goal is to estimate an exposure based metric, which gives more weights to the items towards the top end of the ranking compared to the bottom, it would be better to use a sampling distribution that gives more weights to items towards the top. If uniform sampling, i.e., sampling documents uniformly at random and label them for group membership, were to be used instead, it is likely that we will be spending our annotation budget on items that have little to no impact on the value of the chosen metric.

In this paper, we adopt a sampling strategy proposed by ~\citet{pavlu2007practical}, which was shown to achieve good performance in estimation of top heavy IR metrics such as average precision. The approach is based on using a prior distribution that associates each rank position with a weight. Let $R$ be the length of a given ranked list of items, whose quality we are trying to estimate.  Then, the sampling weight associated with an item that is retrieved at rank $r$ can be computed as:

\begin{equation}\label{eq:weights}
W(r) = \frac{1}{2R}\left(1 + \frac{1}{r}+ \frac{1}{r+1}+...+\frac{1}{R}\right)\approx \frac{1}{2R}\log\left(\frac{R}{r}\right)  
\end{equation}

Typically, we would have many ranked lists (systems), the quality of which needs to be estimated at the same time, using the same sampled dataset. Hence, in order to obtain the final sampling distribution that can work reasonably well across all these systems, we first generate these weights for each items retrieved by each system and then average the weights of each item across all systems, resulting in a single weight for each item. Finally, these weights are converted to a probability distribution by normalizing with the sum of weights over all the items. 

For the actual sampling process, the stratified sampling strategy by Stevens~\cite{brewer2013sampling, stevens1958sampling} that has also been used by Pavlu et al.~\cite{pavlu2007practical} has been shown to have the advantage of achieving reduced variance. Hence, we also adopt this sampling strategy in this paper.  

Let $m$ be the amount of annotation budget we have available. The stratified sampling process works as follows~\cite{pavlu2007practical}:
\begin{enumerate}
\item Order the items in decreasing order based on their sampling probability (Eq.~\ref{eq:weights}), and partition them into buckets of size $m$. 
\item For each bucket, assign a sampling probability for that particular bucket by taking the mean of items' probabilities that fall under each bucket, and normalizing across all buckets. 
\item Sample buckets with replacement $m$ times.
\item Uniformly sample items without replacement from each sampled bucket. If a bucket is sampled $n$ times, sample uniformly $n$ items from that bucket without replacement.
\end{enumerate}

Note that this particular sampling strategy works by first sampling buckets, and then sampling items from each bucket, as opposed to directly sampling items. We call this strategy as \textit{weighted sampling} throughout the paper.

We would like to further emphasize that while we decided to use the sampling distribution described in Equation~\ref{eq:weights} in this work, the estimation technique we use in this paper can work with \emph{any} distribution and produce unbiased estimates. Thus, our approach has the flexibility towards incorporating prior knowledge of group distribution to stakeholders in fairness, i.e. law makers, practitioners such that the sampling distribution can be changed to meet the prior knowledge.  

\subsection{Statistical Estimation of Fairness Metrics}\label{subsec:est}

After obtaining samples drawn from a sampling distribution, we need to derive an estimator to compute the estimated value of a fairness metric. Our approach is based on extending the statistical estimation method from \citet{pavlu2007practical}, which uses the Horvitz-Thompson estimator~\cite{thompsonsamling} for estimating values of IR metrics, to estimating values of fairness metrics. 

\subsubsection{Horvitz-Thompson Estimator for Estimation of the Mean}
 Suppose we are given a sample set $S$ of size $m$ sampled from a population $D$. According to the Horvitz-Thompson estimator~\cite{thompsonsamling}, an unbiased estimate of the population mean $X$ can be computed as:

\begin{equation}\label{eq:horw}
    \widehat{X} = \frac{\sum_{i \in S}\frac{f(\pi_i)}{\theta_i}}{|D|}
\end{equation}

\noindent where $f(\pi_i)$ is the value associated with item $i$ and $\theta_i$ is the \emph{inclusion probability} for item $i$, which represents the probability that item $i$ would be included in \emph{any} sample of size $m$.

One should note that when samples are drawn using the stratified sampling strategy described in the previous section, the inclusion probability of item $i$ is different than the sampling probability associated with this item. Yet, inclusion probabilities can be derived from the sampling probabilities as follows: Let $b$ be the sampling probability for a bucket of size $m$ (where $b$ is the sum of the sampling probabilities of all items that fall under this bucket), and $T$ be a random variable that indicates the number of times this bucket has been selected in the stratified sampling process. Then, $\theta_i$ for an item $i$ within this bucket can be computed as ~\cite{pavlu2007practical}:

\begin{equation}
    \theta_i = \sum_{k=1}^m P(T=k) \frac{k}{m} = \frac{1}{m}E[T] = \frac{1}{m}mb = b
\end{equation}

\subsubsection{Horvitz-Thompson Estimator for Estimation of Fairness Metrics}

Recall that $\treatmentExposure{\group}$ value described in Equation~\ref{eq:disc_sum}, indicates the proportion of group $g$ within the top $k$ ranking results. Then, using Horvitz-Thompson estimator, $\widehat{\treatmentExposure{\group}}$, an unbiased estimator for $\treatmentExposure{\group}$, can be computed as: 

\begin{equation}~\label{eq:frac_est}
    \widehat{\treatmentExposure{\group}} = \frac{1}{k} \sum_{i \in S, rank(i) \leq k} \frac{\indicator{\ranking_i\in\docs{\group}}}{\theta_i}
\end{equation}

Hence, the estimators for the four divergence based fairness metrics defined in Equation~\ref{eq:divergence_metrics} can be computed as:

\begin{align}~\label{eq:estimated_metrics}
   \widehat{\metricDiff} & = \sum_{\group \in \groups} (\targetExposure{\group}  - \widehat{\treatmentExposure{\group}})\\~\label{eq:diff_metric}
   \widehat{\metricAbs} & = \sum_{\group \in \groups} |\targetExposure{\group}  - \widehat{\treatmentExposure{\group}}|\\~\label{eq:square_metric}
    \widehat{\metricSq} & =\sum_{\group \in \groups} \left(\targetExposure{\group}  - \widehat{\treatmentExposure{\group}}\right)^2 \\~\label{eq_kl_metric}
     \widehat{\metricKL} &= \sum_{\group \in \groups} \targetExposure{\group} \log \left(\frac{\targetExposure{\group}}{\widehat{\treatmentExposure{\group}}}\right)
\end{align}

Similarly, the estimator $\widehat{\exposureSum{\group}}$ for group exposure $\exposureSum{\group}$ described in Equation~\ref{eq:exposure_sum} can be computed by substituting the value of each item in the sample with $\gamma^{(rank(i)-1)} \indicator{\ranking_i\in\docs{\group}}$, which leads to the formula below:

\begin{align}~\label{eq:est_exposure}
    \widehat{\exposureSum{\group}} = (1-\patience) \sum_{i \in S, rank(i) \leq k} \patience^{(i-1)} \frac{\indicator{\ranking_i\in\docs{\group}}}{\theta_i}
\end{align}

\section{Experimental Setup}
\label{sec:experiments}

We evaluated the quality of our proposed estimation method across three different experimental conditions.  In this section, we will describe our experimental setup, including datasets, metrics, and baselines.

\subsection{Data}
\subsubsection{Synthetic Data}\label{subsec:synthetic_data} 

We develop a simulator to generate various simulated rankings (systems) in order to analyse the performance of our estimation methods in depth over multiple fair ranking metrics, and we list the variables used in our simulator as follows:

\begin{itemize}
    \item $Q$: number of queries. 
    \item $D$: number of documents in corpus.
    \item $M$: number of systems submitted.
    \item $N$: number of documents retrieved for each query, i.e. $N \leq D$.
    \item $h_q$: parameter modeling the ``easiness'' of query $q$, that is, how difficult the query is for a baseline retrieval system.
    \item $r_{dq}$: simulated relevance of document $d$ for query $q$.

    \item $\alpha_m$: parameter modeling the average effectiveness of system $m$, independent of query.
    \item $\gamma_m$: parameter modeling the bias of a system towards or against a sensitive group.
    \item $\beta$: parameter modeling the distribution of the protected group, $\indicator{\doc \in \docsA}$ defined in Section~\ref{sec:prob}, in corpus.
    \item $\ranking_{\query m}$: system's retrieval result for query $q$.
    \item $\metric(\ranking_{qm})$: fairness performance of a system, in query $q$.
\end{itemize}

The simulator begins by assuming the existence of a test collection with $D$ documents, $Q$ queries, and relevance judgments between every $\langle$ query, document $\rangle$ pair.
These relevance judgments are simulated per query as follows: we first sample, for each of the $Q$ queries, a ``query easiness'' parameter $h_q$ from a prior beta distribution with parameters.
This models the proportion of expected relevant documents for the query $q$.
This parameter is then used in a Bernoulli distribution to sample the binary relevance $r_{dq}$ of each document $d$ to the query $q$.
Via this procedure, we obtain a simulated set of queries with varying numbers of relevant documents, some much more than others, which is typical of an information retrieval test collection.

We next simulate protected class.
We make the assumption that protected class is independent of relevance, and independent of query; we assume it is a global property of a document.
Parameter $\beta$ models the expected proportion of protected group members.
Similarly to relevance, protected group status is sampled from a Bernoulli distribution with parameter $\beta$.

Retrieval systems are then simulated by sampling document scores $S_{dq}$ and ranking them in decreasing order.
A score $S_{dq}$ is a function of the query easiness parameter $h_q$ described above, a ``system goodness'' parameter $\alpha_m$, the relevance $r_{dq}$, the protected group status, and a global ``group bias'' parameter $\gamma$.
Specifically, the score is sampled as follows, with different cases for different combinations of relevance $r_{dq}$ and protected group membership $I(d \in \mathcal{D}_A)$:
\begin{align*}
    S_{dq} \sim \begin{cases}
      \mathcal{N}\left(0, \sigma\right) & \text{if $r_{dq} = 0$ and $I(d \in \mathcal{D}_A) = 0$} \\
      \mathcal{N}\left(\alpha_m + h_q, \sigma\right) & \text{if $r_{dq} = 1$ and $I(d \in \mathcal{D}_A) = 0$} \\
      \mathcal{N}\left(\gamma, \sigma\right) & \text{if $r_{dq} = 0$ and $I(d \in \mathcal{D}_A) = 1$} \\
      \mathcal{N}\left(\alpha_m + h_q + \gamma, \sigma\right) & \text{if $r_{dq} = 1$ and $I(d \in \mathcal{D}_A) = 1$} 
    \end{cases}
\end{align*}
In effect, this process results in relevant documents having higher scores correlated to both system goodness and query easiness, and protected group members having higher scores correlated to global bias.
Once scores have been generated for all documents for a query $q$, they are ranked in decreasing order by score ($\pi_{qm}$), and then all standard IR and fairness measures can be computed over the ranking.

We can simulate a wide variety of different experimental conditions by manipulating the variables $D$, $Q$, $M$ and parameters for prior beta distributions and group membership proportion. By setting the $\beta$ parameter for protected group membership to $(1,1)$, we made sure that both groups have equal presence in the generated corpus. Hence, a fairness target of $0.5$ is being used for divergence based metrics described in Section~\ref{subsec:disc_met}. 

In our experiments, we generated $M=800$ systems retrieving $N=100$ documents from a corpus containing $D=1000$ documents for $Q=50$ queries.

\begin{table}[!ht]
\label{tab:synth_data}
\begin{tabular}{lr}
\toprule
Number of  Systems (M)                & 800   \\ 
Number of Queries (Q)                 & 50  \\
Number of Documents (D)               & 1000 \\
Number of Retrieved Documents (N)     & 100 \\
Average documents in protected group per query & $\approx$50\\
\bottomrule
\end{tabular}
\caption{Summary of Synthetic Data.}
\end{table}

\subsubsection{Real-World Data}
Beyond the synthetic data, we test the performance of our method using two real-world datasets: (1) TREC Fair Ranking dataset consisting of submissions to the TREC Fair Ranking Track~\cite{trec-fair-ranking-2019}, and (2) book recommendation data from \citet{ekstrand:bookdata} consisting of ranked lists of books recommendations. Below we describe each dataset in more detail:

\paragraph{TREC Fair Ranking Dataset}\label{subsec:trec_data}
In the TREC Fair Ranking Track, each participant was asked to re-rank a given initial list of documents for each query. This dataset contains the binary protected attributes of group membership for each of document, as well as relevance judgments for each query-document pair.  Since each participant of the Fair Ranking Track was given the same initial list of documents to re-rank, the proportion of each group is identical for all submitted systems~\cite{biega2020overview}. Hence, divergence based metrics that depend on group proportions are not directly applicable to this dataset. Instead, Exposure based fairness metrics (explained in Equation~\ref{eq:exposure_sum}) can be used to compare fairness among different systems. We use exposure of group called \textit{advanced} in the annotations to compute our Exposure metric. Details about this dataset can be seen in the table below.

\begin{table}[!ht]
\label{tab:fair_trec}
\begin{tabular}{lr}
\toprule
Number of  Systems (M)                & 10   \\ 
Number of Queries (Q)                 & 635  \\
Number of Documents (D)               & 2671 \\
Average number of documents per query & 7    \\
Mode of documents per query           & 6    \\
Average member from protected group per query & $\approx$4\\
\bottomrule
\end{tabular}
\caption{Summary of TREC Data.}
\end{table}

\paragraph{Book Recommendation}
\label{sec:exp:book-recs}
We also test our estimation method on the book data and recommendation models assembled by \citet{ekstrand:bookdata}.
This dataset combines user-book interaction data from GoodReads \citep{Wan2018-monotonic} with book metadata from the Library of Congress and OpenLibrary. For each book, the dataset also contains a binary atrribute dictating the gender of the author\footnote{Binary gender is a limitation of the underlying author metadata; see \citet{ekstrand:bookdata} for a detailed discussion of data limitations.}.
From this data, we generated 100-item recommendation lists for 5000 users via matrix factorization method using implicit feedback~\cite{Takacs2011-ix}.
The dataset contains a ranked list of book recommendations for each user. Hence, the fairness metrics can be computed for each user, treating women as the protected group~\cite{ekstrand:bookdata}. The gender distribution in the pooled corpus of \emph{all} recommended books is used as our fairness target for divergence based metrics. Table~\ref{tab:my_label} shows more details about this dataset.

\begin{table}[!ht]
    \begin{tabular}{lr}
\toprule
Number of Users (M)                   & 5000   \\ 
Number of Books (D)                   & $\approx$500K  \\ 
Average number of books per user      & 100    \\ 
Average books from protected group per user & $\approx$24\\
\bottomrule
\end{tabular}

    \caption{Summary of Book Recommendation Data.}
    \label{tab:my_label}
\end{table}

\subsection{Sampling Setup}

We simulate the setup of not having complete annotations available by sampling from the set of complete judgments using different sampling rates $p \in \{0.1, 0.2, 0.3, 0.4, ..., 0.9\}$, where sampling using a sampling rate of $p=0.1$ corresponds to generating a dataset that contains $10\%$ of the complete judgements.

The simulated dataset and the TREC Fair dataset contains two types of annotations: the relevance judgements associated with each query document pair, as well as the protected attribute associated with each document.  In our experiments, we assume that complete relevance judgments are available, and that only annotations related to the protected attributes are incomplete. Hence, when we generate our sampled datasets, we only sample from the protected attribute annotations. 

For the Book Recommendation dataset, we assume that all recommended items are relevant and sample from the gender attribute.

\subsection{Comparison of Estimated vs. Actual Values}

Given a sampled dataset, we  compute the estimated values of the various fairness metrics using our proposed estimators for each system. We then evaluate the quality of our estimations by comparing them with actual metric values (computed using all the available judgments, as opposed to just the sampled ones) using the following statistics: 
\begin{enumerate}[label=(\roman*)]
    \item \textit{Kendall's $\tau$}: This statistic is used to compute the correlation between two system rankings.  Its value can range from $-1$ (perfectly negative rank correlation)  to $1$ (perfectly correlated).
    \item \textit{Root Mean Squared Error ($RMSE$)}: This statistic is used to compare the actual and estimated metric values across all the systems in the experiment.
\end{enumerate}

\section{Baselines}
\label{sec:baselines}

We compare are method against two baseline approaches, a non-sampling technique and uniform sampling.

\subsection{Induced Method Baseline}

We first compare against \textit{induced metrics} proposed by \citet{yilmaz2006estimating}, which are shown to achieve reasonable performance for estimating IR metrics when judgments are incomplete.  The induced version of a metric is computed by filtering the ranking to only include the labelled items instead of measuring the whole ranking based on the sample. This is done by removing the unlabelled items from the list, as a result of which lower-ranked labelled items move up in the ranking. The metric is then computed over this induced ranking that only contains labelled items.

\subsection{Uniform Sampling Baseline}
In a uniform sampling setup, instead of using our top heavy sampling distribution, sampling is performed uniformly at random, all items having equal likelihood of being included in the sample.

When uniform sampling is used in the sampling phase, estimation becomes quite straightforward as the actual mean can simply be estimated by computing the simple sample mean. Hence, in such a setup the estimator for $\treatmentExposure{\group}$ can be computed as:

\begin{equation}~\label{eq:frac_uniform_est}
    \widehat{\treatmentExposure{\group}} = \frac{1}{k} \sum_{i \in S, rank(i) \leq k} \indicator{\ranking_i\in\docs{\group}}
\end{equation}

Substituting $\widehat{\treatmentExposure{\group}}$ in the set of equations from Equation~\ref{eq:estimated_metrics} to Equation~\ref{eq_kl_metric} corresponds to the estimations of various divergence based fairness metrics under uniform sampling.

Similarly, the estimator $\widehat{\exposureSum{\group}}$ for the Exposure metric can be computed as: 

\begin{align}~\label{eq:est_uniform_exposure}
    \widehat{\exposureSum{\group}} = (1-\patience)\frac{1}{k} \sum_{i \in S, rank(i) \leq k} \patience^{(i-1)} \indicator{\ranking_i\in\docs{\group}}
\end{align}

\section{Results}
\label{sec:results}

In this section, we examine our method's performance under different experimental conditions. We mainly focus on the scenario when $10\%$ of the complete annotations are available (corresponding to sampling rate $p = 0.1$), although we also report aggregate results for different sampling rates. 

In what follows, we first compare our fair ranking metric estimations against the induced method on both the synthetic and the real-world datasets. We then compare the quality of estimators obtained using weighted sampling with that of uniform sampling.

\subsection{Estimated vs. Induced Metrics}
We test how our estimation method performs against the \emph{induced} method using both the synthetic and the real world-data. 

\begin{figure}[!ht]
    \begin{subfigure}[b]{0.45\textwidth}
    \includegraphics[trim={1cm 1cm 1cm 1cm},clip,width=8cm]{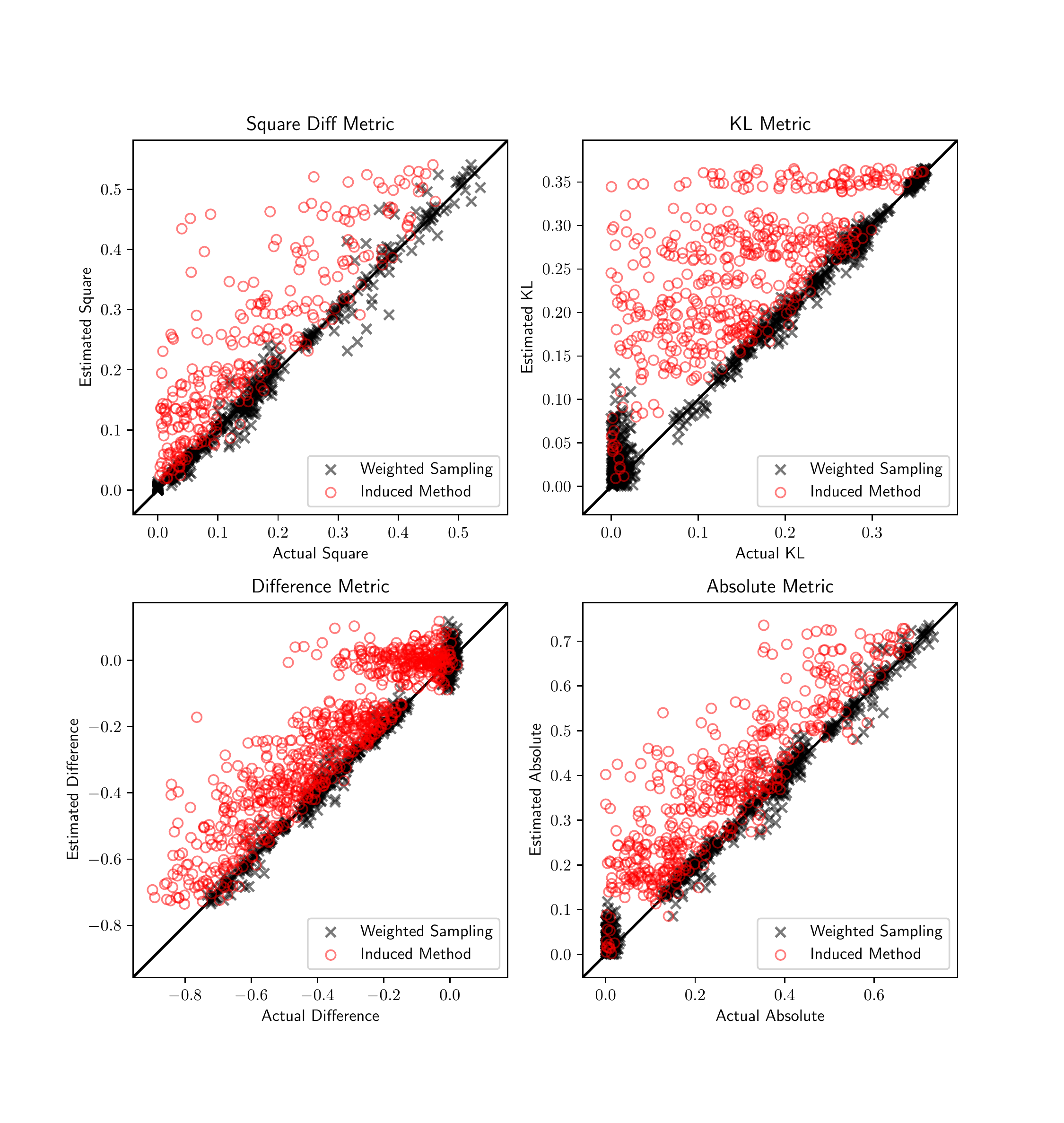}
    \caption{Synthetic Data}\label{fig:metric_results}
    \end{subfigure}\hfill
    \begin{subfigure}[b]{0.45\textwidth}
    \includegraphics[trim={1cm 1cm 1cm 1cm},clip,width=8cm]{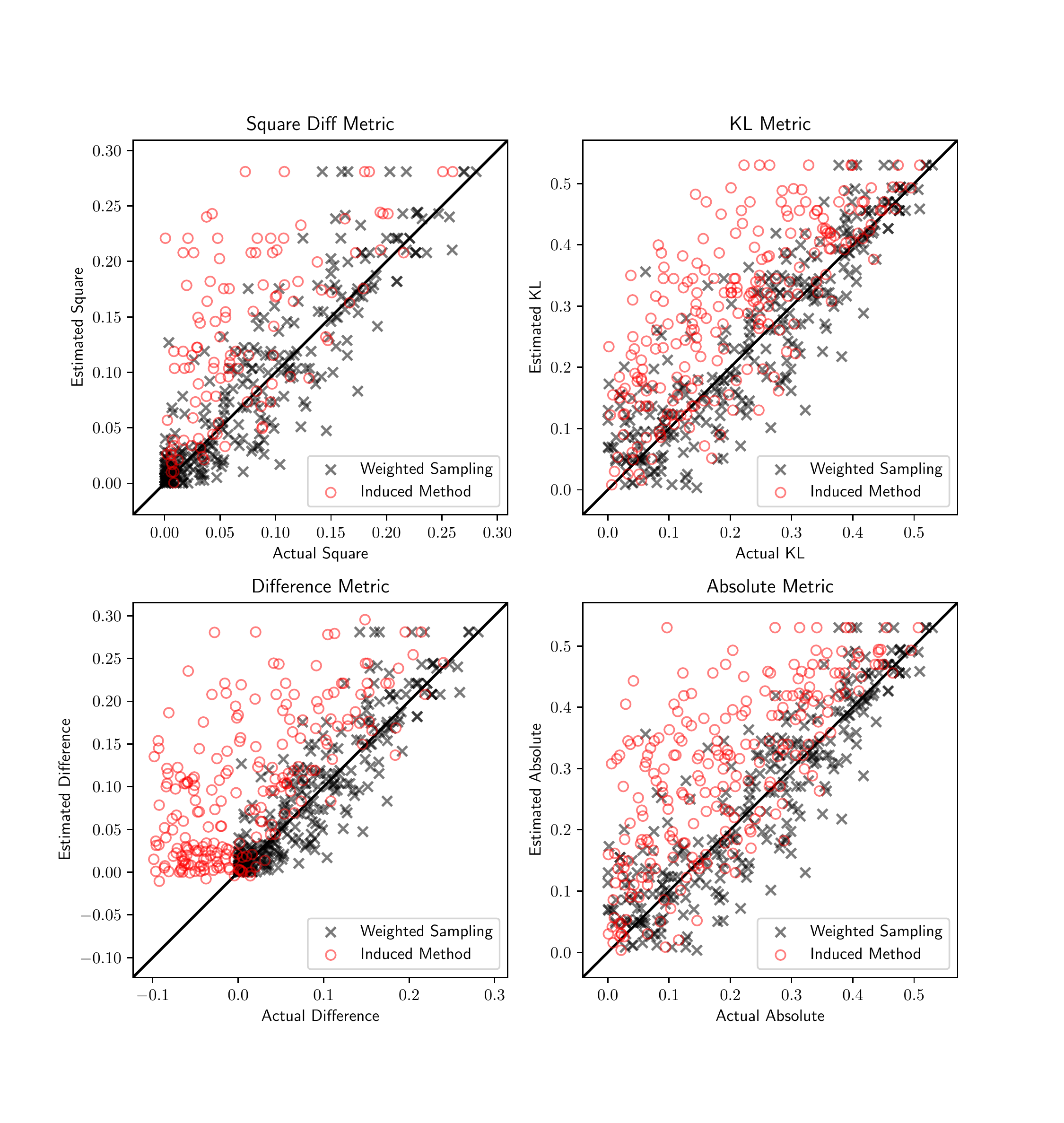}
    \caption{Book Data}\label{fig:recom_metric_results}
    \end{subfigure}

    \caption{Estimation of divergence based fairness metrics using the proposed method versus the induced baseline. Each dot in the synthetic data represents the mean performance of each system over different queries; in book recommendation data, each dot represents the performance of recommendations per user. } %
\end{figure}

Figures~\ref{fig:metric_results} and~\ref{fig:recom_metric_results} depict our results for the four divergence based fair ranking metrics for the synthetic and the book recommendation datasets, respectively. The $x$ axis in these figures show the actual metric value (if we had complete judgments available), while the $y$ axis shows the estimated values computed using incomplete judgments. For comparison purposes, we also plot the line $y=x$ in the plots. As it can be seen, our estimates are generally well-distributed around the line $y=x$, validating that they are unbiased. On the other hand, the induced baseline tends to over-estimate the actual values consistently for both datasets. Since the proportions of different groups are identical in each system's output in the TREC Fair dataset, we do not report any proportion based metric results for this dataset.

\begin{figure}[htb]
    \begin{subfigure}[b]{0.40\textwidth}
    \includegraphics[width=8cm]{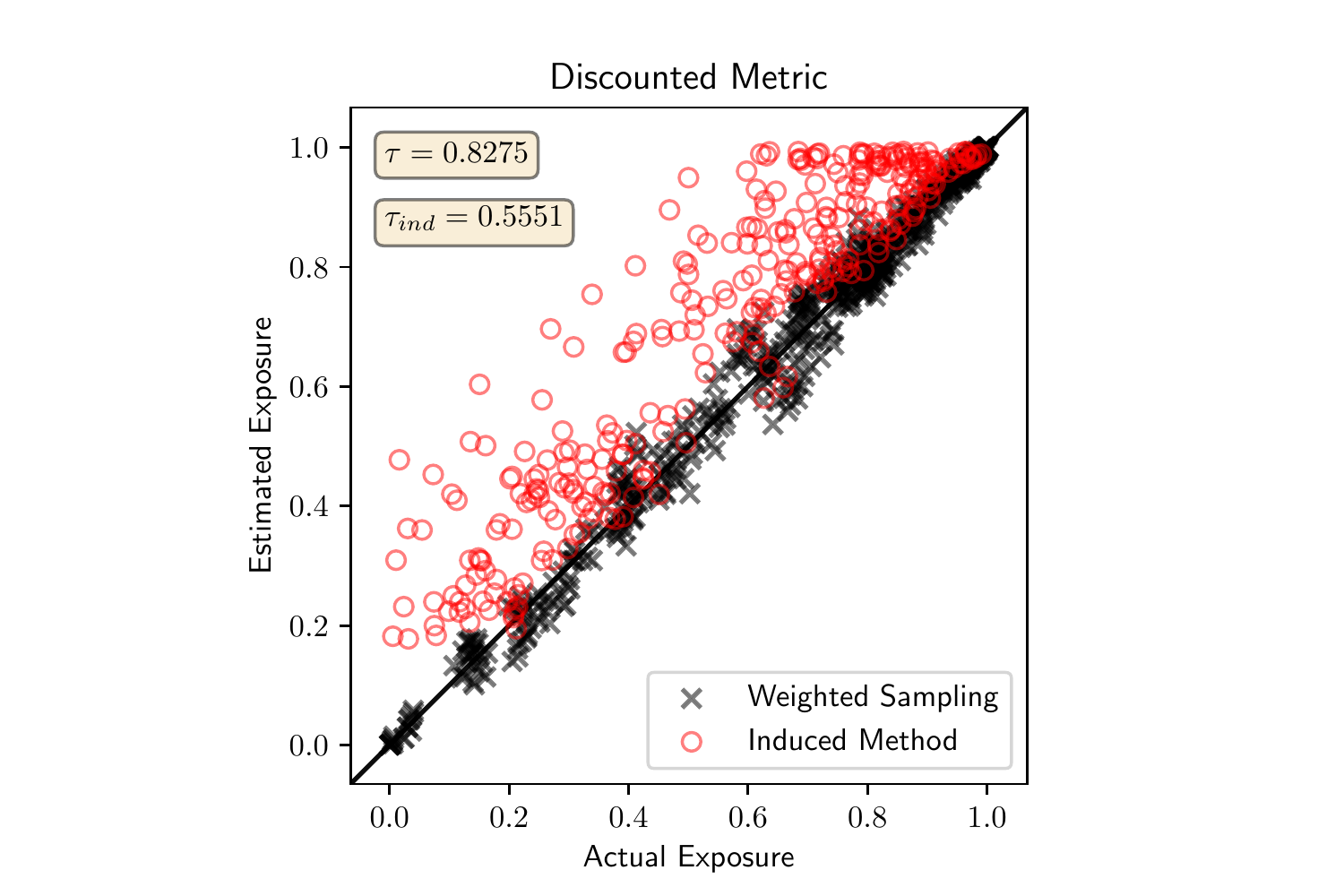}
    \caption{Synthetic Data}\label{fig:ind_samp}
    \end{subfigure}
    
    \begin{subfigure}[b]{0.40\textwidth}
    \includegraphics[width=8cm]{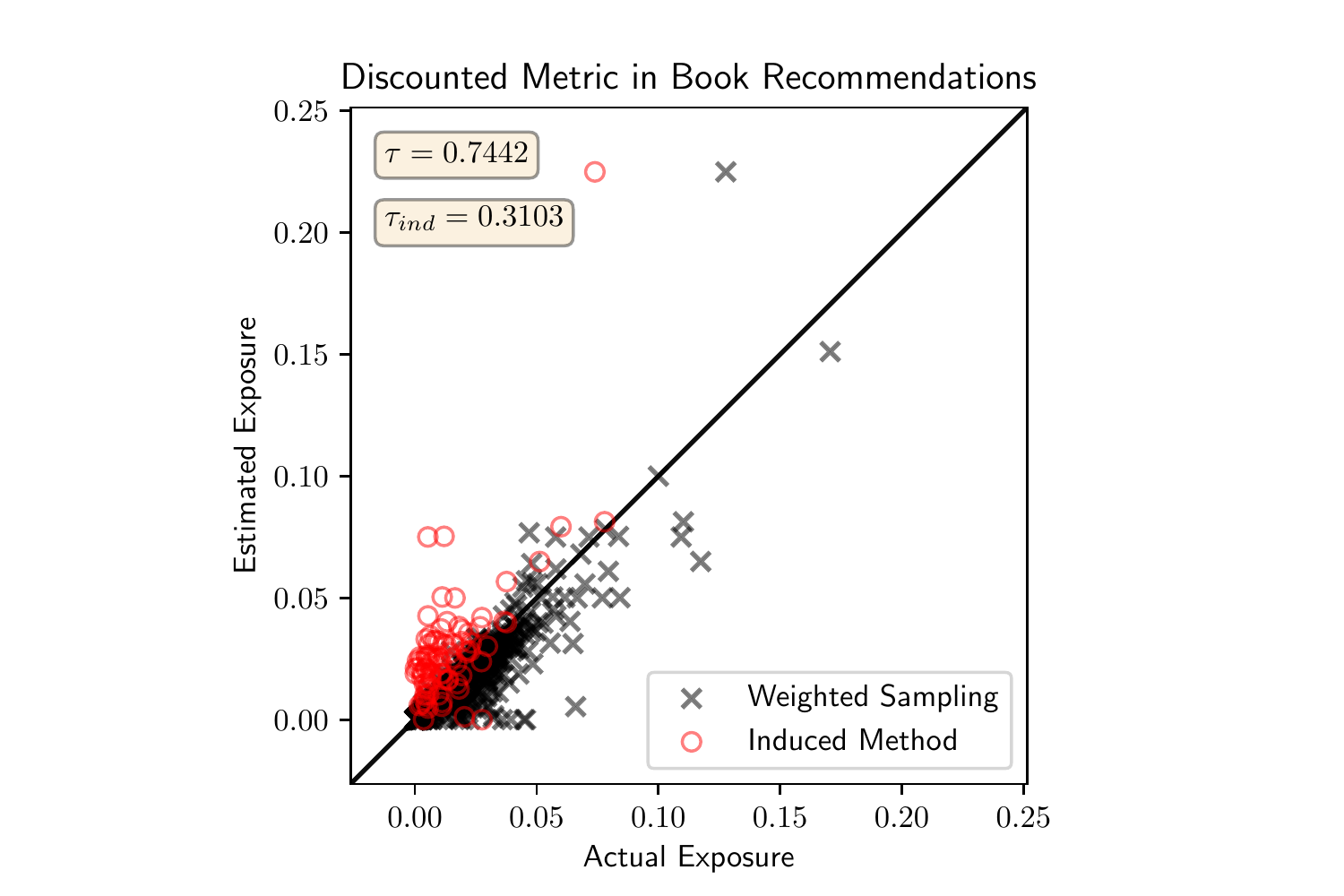}
    \caption{Book Data}\label{fig:recom_disc}
    \end{subfigure}
    \caption{Estimation of the exposure metric. $\tau$ and $\tau_{ind}$ denote the Kendall's $\tau$ correlation coefficients for weighted and induced methods respectively. Each dot in synthetic data represents the mean performance of each system over different queries; in book recommendation data, each dot represents the performance of recommendations per user. The horizontal axis denotes the actual value for exposure metric and vertical axis is the estimated/induced values in both figures.}
\end{figure}

Next, we examine the estimation of the discounted metrics on the synthetic and the book recommendation datasets, results of which can be seen in Figures~\ref{fig:ind_samp} and~\ref{fig:recom_disc}. The plots show the Kendall's $\tau$ correlations between the actual and estimated values. 

 \begin{table}[tb]
\caption{RMSE and Kendall's $\tau$ values for the proposed vs. induced method in synthetic data.}
\label{tab:results_we_ind}
\begin{tabular}{R{1.5cm}|C{0.7cm}C{0.7cm}|C{0.7cm}C{0.7cm}}
\toprule
    & \multicolumn{2}{c|}{Weighted} & \multicolumn{2}{c}{Induced}\\
      & RMSE    & $\tau$  & RMSE    & $\tau$ \\ \hline
$\metricAbs$    & 0.0332 & 0.8112 & 0.1361 & 0.5792\\
$\metricSq$     & 0.0303 & 0.8014 & 0.1103 & 0.5978\\
$\metricKL$     & 0.0298 & 0.8413 & 0.1131 & 0.5725\\
Exposure      & 0.0341 & 0.8275 & 0.1412 & 0.5551\\
\bottomrule
\end{tabular}

\end{table}

Note that all the previous reported results are over a \emph{single} sampled dataset, which could affect the conclusions reached due to the randomness in the sampling process. In order to make sure our results are not affected by the particular sample chosen in the sampling phase, we further generate $10$ different samples using a sampling rate of $p = 0.1$ and report the mean RMSE and mean Kendall's $\tau$ values across these $10$ samples.  Table~\ref{tab:results_we_ind} demonstrates the RMSE and Kendall's $\tau$ values when comparing actual and estimated metric values for the various divergence based metrics and the Exposure metric using the synthetic dataset.  As it can be seen, the proposed method results in much lower RMSE values and higher Kendall's $\tau$ correlations when compared to the induced method.

\begin{figure}[tb] 
    \begin{subfigure}[b]{0.40\textwidth}
    \includegraphics[width=8cm]{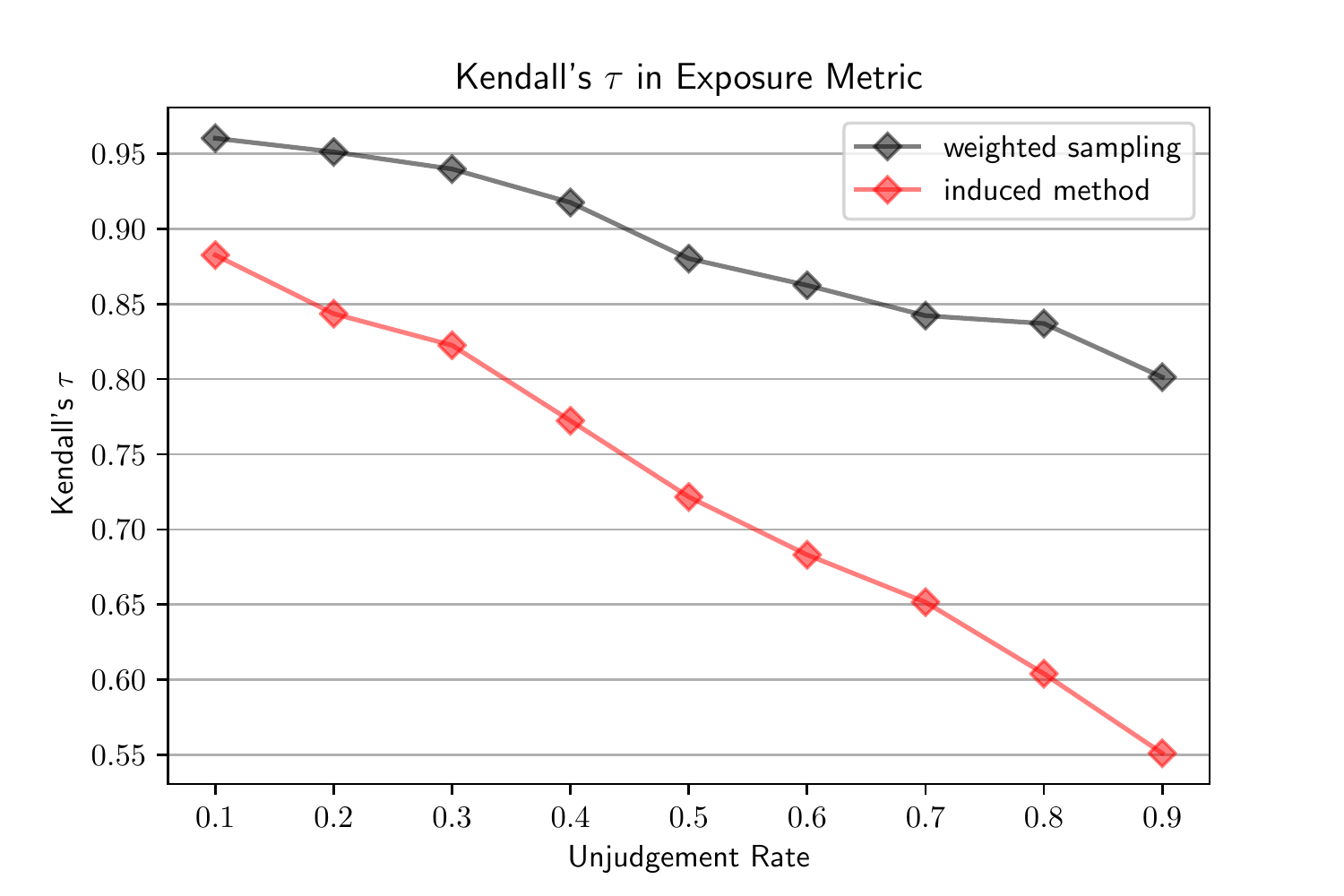} %
    \caption{Protected Group Exposure Metric}\label{fig:tau_comp_wind}
    \end{subfigure}
    
    \begin{subfigure}[a]{0.40\textwidth}
    \includegraphics[width=8cm]{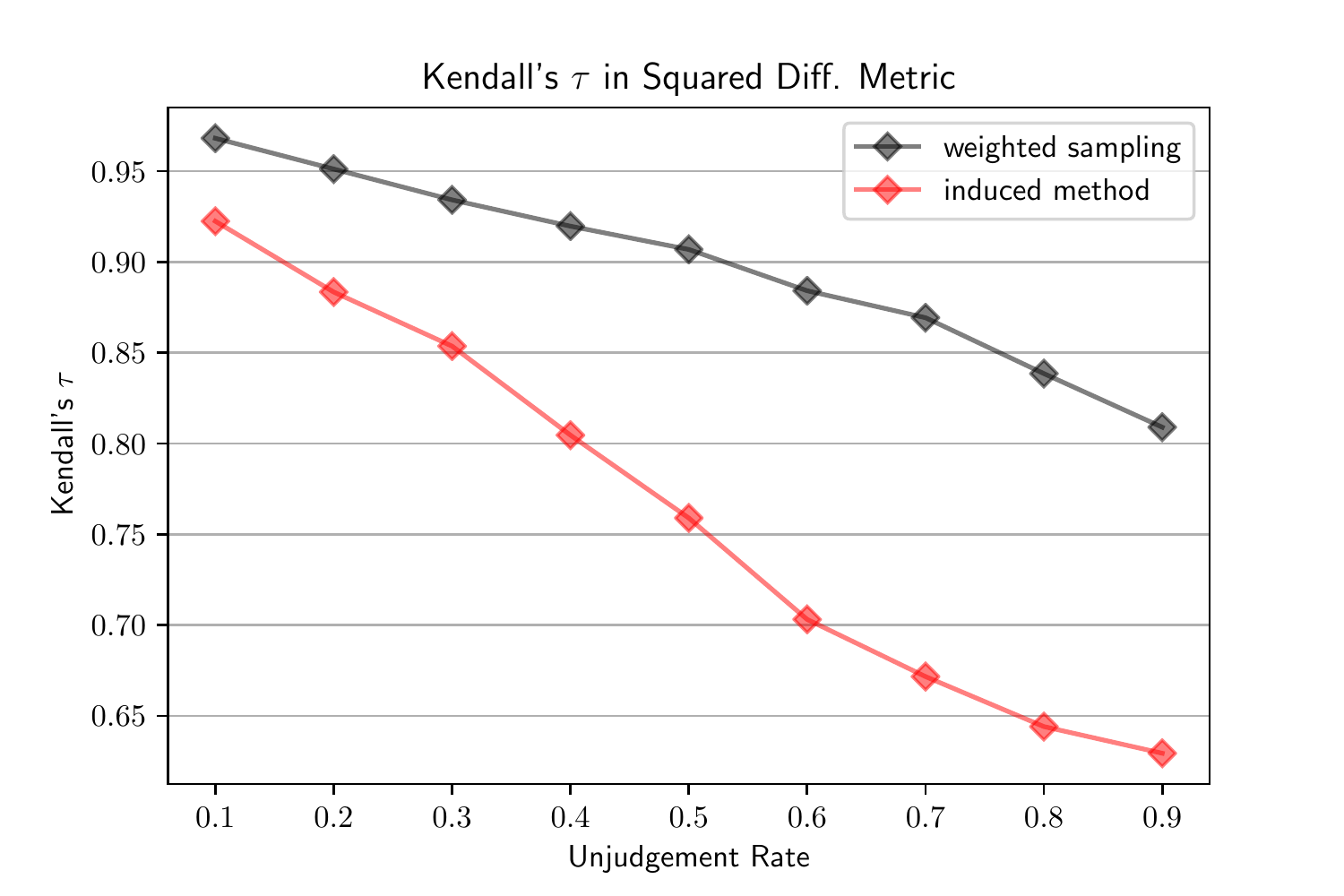}
    \caption{Squared Difference Proportion Metric}\label{fig:sq_tau_comp_wind}
    \end{subfigure}\hfill
    \caption{Comparison of weighted sampling and induced method for (top) Protected Group Exposure metric and (bottom) Squared Difference metric.} %
\end{figure}

While the results we have presented until now mainly focus on a sampling rate $p = 0.1$, our results seem consistent across different sampling rates. Figure~\ref{fig:tau_comp_wind} and Figure~\ref{fig:sq_tau_comp_wind} show how the quality of our proposed method compares with that of induced method for various sampling rates $p \in \{0.1, 0.2, .... 0.9\}$. In these plots, we focus on the Squared Difference metric and the Exposure metric\footnote{We observed similar results for other divergence metrics.} as the evaluation metrics. The $x$ axis in the plots show the rate of unjudged documents ($1-p$) when sampled datasets are generated using various sampling rates $p$, and the $y$ axis shows the Kendall's $tau$ correlations between the actual vs. estimated values. In order to ensure that the results are not affected by a particular sample, the reported values for each sampling rate are the average Kendall's $\tau$ correlation values across $10$ different randomly sampled datasets. It can be seen that the estimation method consistently outperforms the induced method over all the different sampling  percentages, with the gap between the  two methods increasing as judgments become more incomplete. 

Overall, our results show that the proposed statistical estimation technique together with the weighted sampling strategy results in more accurate estimates of fairness metrics compared to the induced method when judgments are incomplete.

\subsection{Weighted vs. Uniform Sampling}\label{Sec:UniformSampling}

Having shown that our weighted sampling strategy outperforms the induced method, we now use the TREC Fair Ranking and the synthetic datasets (Section~\ref{subsec:trec_data}) to show how the estimators based on our proposed method using weighted sampling compare with estimators based on uniform sampling. In this section, we use the formulas based on a simple mean estimator (as described in Section~\ref{Sec:UniformSampling}) when uniform sampling is used.

\begin{figure}[!h]
    \begin{subfigure}[b]{0.45\textwidth}
   \includegraphics[trim={2cm 0cm 1.3cm 0cm}, clip, width=8cm]{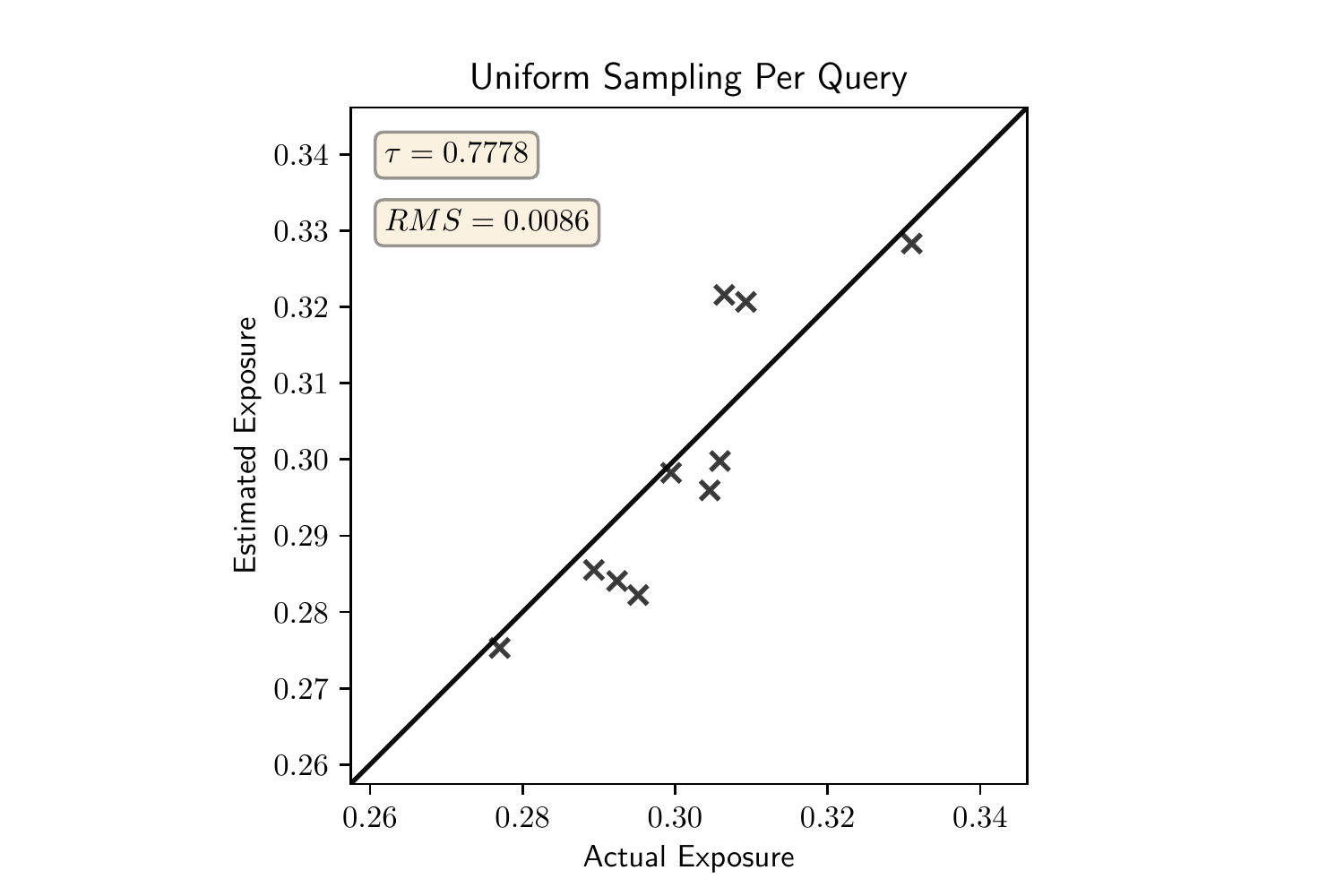}
    \caption{}\label{fig:trec_results_rel3_a}
    \end{subfigure}\hfill
    \begin{subfigure}[b]{0.45\textwidth}
    \includegraphics[trim={2cm 0cm 1.3cm 0cm}, clip, width=8cm]{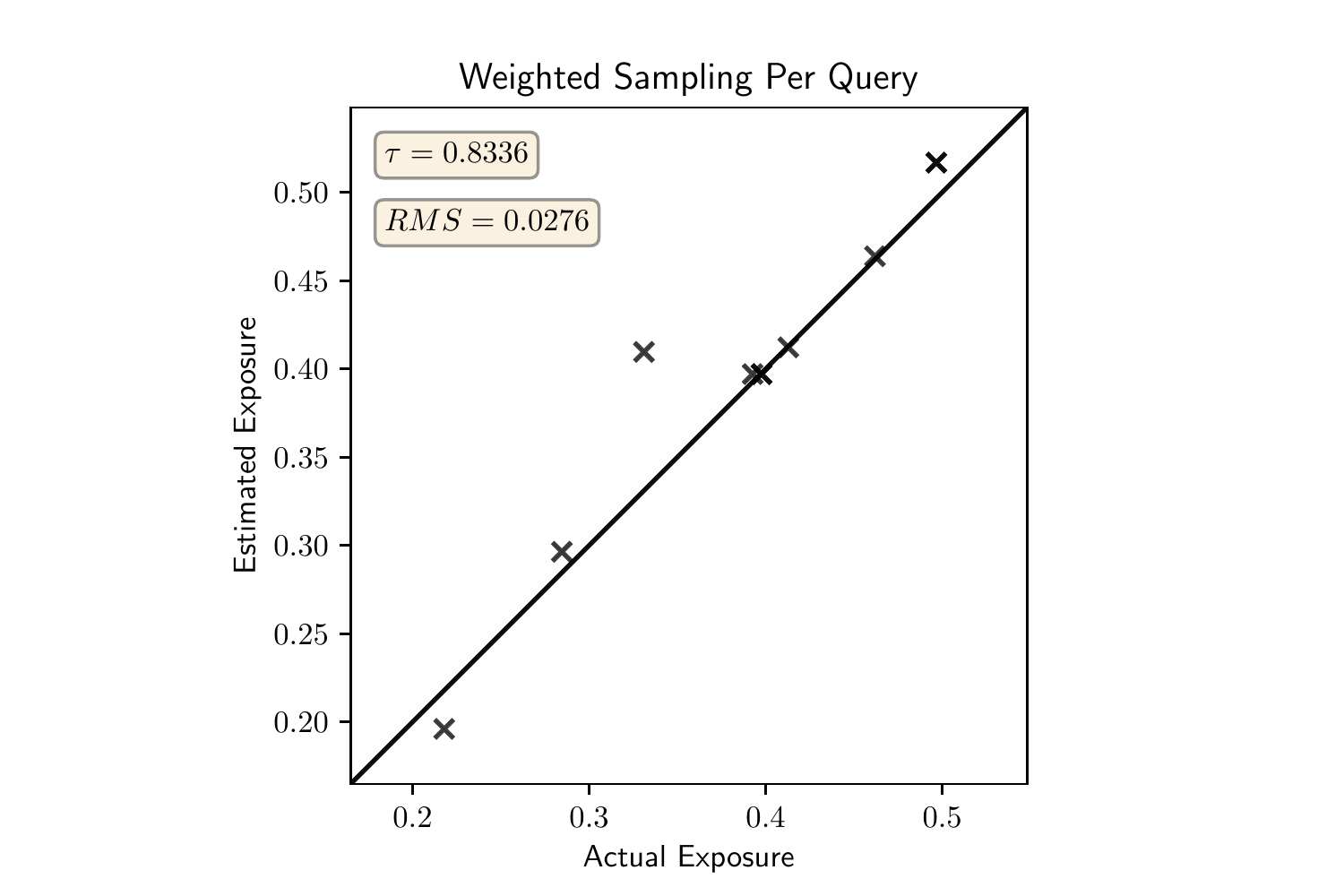}
    \caption{}\label{fig:trec_results_rel3_b}
    \end{subfigure}
    \caption{Estimations on TREC Fair Ranking data for the Protected Group Exposure metric.}
\end{figure}

Figure~\ref{fig:trec_results_rel3_a} and Figure~\ref{fig:trec_results_rel3_b} show the quality of estimations with a sampling rate of $p = 0.1$ for the protected group exposure metric on the TREC dataset. The proposed estimator based on weighted sampling results in much better estimates compared to uniform sampling.

 \begin{table}[!h]
\caption{RMSE and Kendall's $\tau$ values for weighted vs. uniform sampling in synthetic data.}
\label{tab:results_we_uni}
\begin{tabular}{R{1.5cm}|C{0.7cm}C{0.7cm}|C{0.7cm}C{0.7cm}}
\toprule
    & \multicolumn{2}{c|}{Weighted} & \multicolumn{2}{c}{Uniform}\\
      & RMSE    & $\tau$  & RMSE    & $\tau$ \\ \hline
$\metricAbs$    & 0.0332 & 0.8112 & 0.0253 & 0.8013\\
$\metricSq$     & 0.0303 & 0.8014 & 0.0348 & 0.8045\\
$\metricKL$     & 0.0298 & 0.8413 & 0.0289 & 0.7981\\
Exposure      & 0.0341 & 0.8275 & 0.0421 & 0.7147\\
\bottomrule
\end{tabular}
\end{table}

\begin{figure}[!ht] 
     \begin{subfigure}[b]{0.40\textwidth}
    \includegraphics[width=8cm]{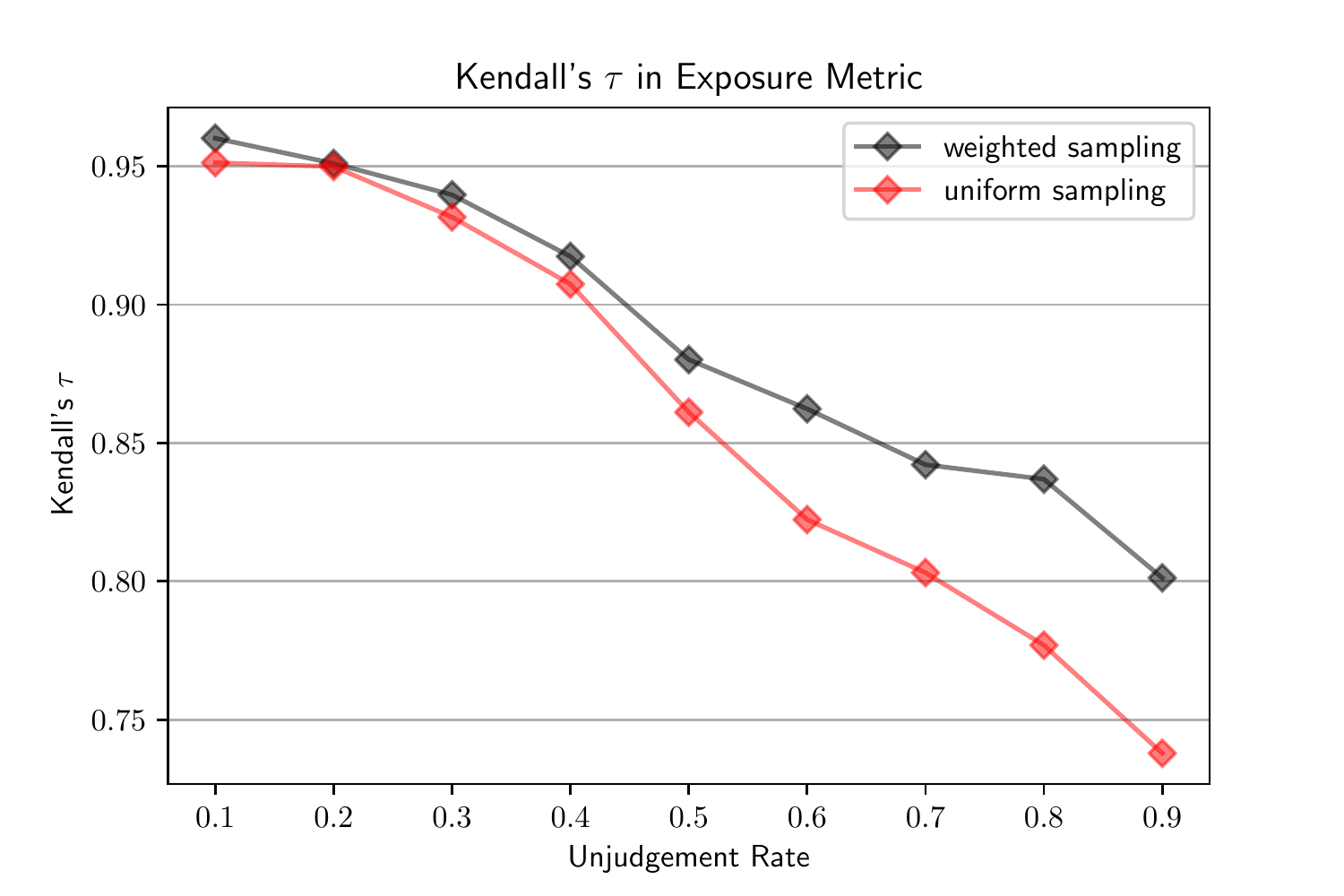} %
    \caption{Protected Group Exposure Metric}\label{fig:tau_comp_wuni}
    \end{subfigure}
    \begin{subfigure}[a]{0.40\textwidth}
    \includegraphics[width=8cm]{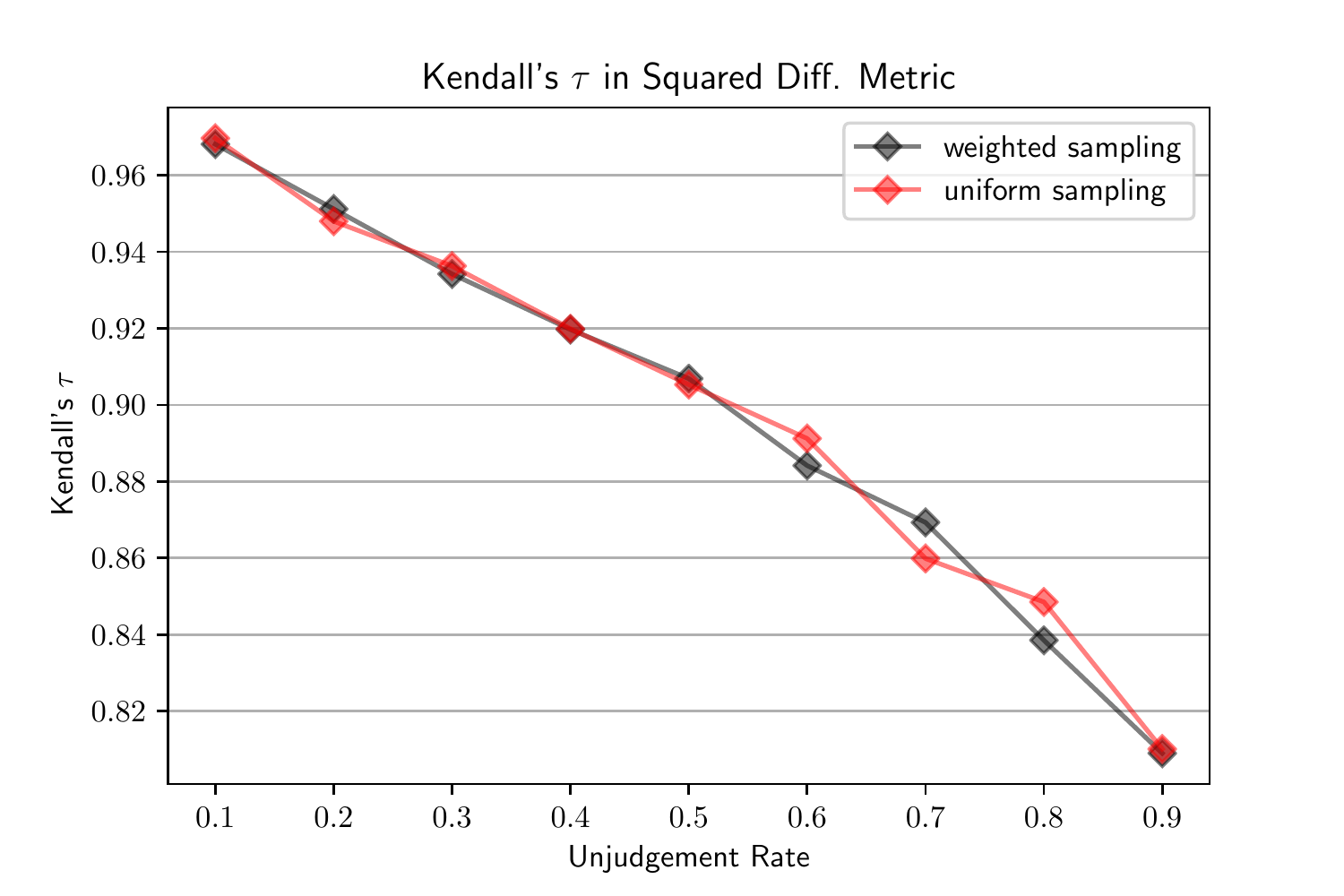}
    \caption{Squared Difference Proportion Metric}\label{fig:sq_tau_comp_wuni}
    \end{subfigure}\hfill
    \caption{Comparison of different estimation techniques for(top) Protected Group Exposure  metric and (bottom) Squared Difference metric.} %
\end{figure}

Table~\ref{tab:results_we_uni} demonstrates the RMSE and Kendall's $\tau$ values when comparing the actual and the estimated metric values for the various proportion based metrics and the Protected Group Exposure metric using the simulated dataset when weighted vs uniform sampling is used. Similar to the setup in Table~\ref{tab:results_we_ind}, in order to make sure our results are not affected by the particular sample chosen, for this experiment, we generate $10$ different samples using a sampling rate of $p = 0.1$ and report the mean RMSE and mean Kendall's $\tau$ values across these $10$ samples. In order to further compare how the quality of the two estimators change across different sampling rates, Figure~\ref{fig:tau_comp_wuni} and Figure~\ref{fig:sq_tau_comp_wuni} show how the weighted sampling estimates compare with that of using uniform sampling for the Protected Group Exposure metric and Squared Difference metric, for various sampling rates $p$. 

Since Protected Group Exposure is a top-heavy metric, giving more weight to the items towards the top end of a ranking, the estimates obtained using our proposed method, which also uses a top-heavy sampling distribution gives much better results compared to uniform sampling.  On the other hand, for the estimation of divergence based metrics such as Squared Difference Metric, which gives equal weight to all items in the ranking, our proposed method performs very similar to uniform sampling. This result further validates that our proposed method produces unbiased estimates of metrics even for metrics that are not top-heavy, even though the sampling distribution used in the sampling process is a top-heavy one.

\section{Conclusion and Future Work}
\label{sec:conclusion}

In this paper, we have adopted the statistical estimation framework developed by Pavlu et al.~\cite{pavlu2007practical} to estimating values of various fair ranking metrics in a scenario where protected group annotations are incomplete. In particular, we
developed techniques based on the Horvitz-Thompson estimator~\cite{thompsonsamling} for estimating five different fairness metrics that fall under two classes of fairness notions---proportion-based and exposure-based.

Our results show that the proposed method, which uses a top heavy sampling strategy, results in unbiased estimates of fair ranking metrics and outperforms naive baselines such as the induced baseline by Yilmaz et al.~\cite{yilmaz2006estimating}, as well as uniform random sampling. 

For future work, we aim to explore three directions. First, although focusing on binary protected attributes covers many common scenarios, expanding our estimation technique to more than two groups and multiple protected attributes is still an important aspect for intersectional fairness issues. Second, we plan to further extend our statistical estimation method towards the estimation of new fairness metrics. Last but not least, we plan to work on identifying sampling distributions that would minimize the number of annotations needed, while achieving high confidence estimates.

\begin{acks}
This project was funded by the EPSRC Fellowship titled ``Task Based Information Retrieval'', grant reference number EP/P024289/1. Michael Ekstrand’s contribution to this work was supported by the National Science Foundation under Grant No. IIS 17-51278.
\end{acks}


\bibliographystyle{ACM-Reference-Format}
\bibliography{main}

\end{document}